\documentstyle[tighten,preprint,pra,aps]{revtex}
\newcommand{\eref}[1]{(\ref{#1})}
\begin{document}

\draft
\title{Electric dipole moment of the electron in YbF molecule.}
\author{N.\ S.\ Mosyagin, M.\ G.\ Kozlov, and A.\ V.\ Titov}
\address{Petersburg Nuclear Physics Institute, \\
         Gatchina, St.-Petersburg district, 188350, RUSSIA}
\date{\today}
\maketitle

\begin{abstract}
{\it Ab initio} calculation of the hyperfine,
$P$-odd, and $P,T$-odd constants for the YbF molecule was performed
with the help of the recently developed technique, which allows to take
into account correlations and polarization in the outercore region.  The
ground state ($^2\Sigma_{1/2}$) electronic wave function of the YbF
molecule is found with the help of the Relativistic Effective Core
Potential method followed by the restoration of molecular four-component
spinors in the core region of ytterbium in the framework of a
non-variational procedure. Core polarization effects are included with the
help of the atomic Many Body Perturbation Theory for Yb atom. For the
isotropic hyperfine constant $A$, accuracy of our calculation is about 3\%
as compared to the experimental datum.
The dipole constant $A_{\rm d}$ (which is much smaller in magnitude), though
better than in all previous calculations, is still underestimated by
almost 23\%.
    Being corrected within a semiempirical approach for a perturbation of
    $4f$-shell in the core of Yb due to the bond making, this error is
    reduced to 8\%.
Our value for the effective electric field on the unpaired electron is 4.9
a.u.=$2.5 \times 10^{10}$ V~cm$^{-1}$.

\end{abstract}



\paragraph{Introduction.}

A number of papers~\cite{Kozlov,Titov,Kozlov1,Quiney,Parpia} were devoted
to the calculations of the $P,T$-odd interaction constants in the ground
state $^2\Sigma_{1/2}$ of YbF molecule. These calculations are necessary
for the interpretation of the ongoing experimental search for the Electric
Dipole Moment (EDM) of the electron $d_{\rm e}$~\cite{Hinds}. It is
expected that in this experiment it will be possible to put a more
stringent bound on the EDM of the electron. However, in order to link
experimentally measured $P,T$-odd frequency shift with $d_{\rm e}$, one
needs to know an effective electric field on the uncoupled electron, which
is characterized by a constant $W_d$. So, reliable calculations of this
quantity are essential. The cited above calculations predict the values of
this constant in the interval
  $(-0.62 \div -1.5) \times 10^{25}\mbox{ Hz e}^{-1}\mbox{ cm}^{-1}$.

It is known, that parameter $W_d$ is sensitive to the spin density in the
vicinity of the heavy nucleus (see, for example,~\cite{KL}).
    The same, of course, can be said about magnetic hyperfine constants.
    So, comparison of the calculated constants of the hyperfine structure
    on the $^{171}$Yb nucleus with the experiment provides an important test
    of the quality of the $W_d$ calculation.
In previous calculations
\cite{Titov,Quiney} hyperfine constants were significantly underestimated.
In~\cite{Titov}, it was concluded that the spin-correlation effects of the
unpaired electron with the outermost core $5s$- and $5p$-shells of
ytterbium should be taken into account in order to perform more accurate
calculations of the hyperfine and $P,T$-odd constants. After that it was
shown that $4f$-shell can also contribute to the dipole part of the spin
density~\cite{Kozlov1}.
    In unrestricted Dirac-Fock (DF) calculations by Parpia~\cite{Parpia}, the values for
    $W_d$ are in a reasonable (mutually consistent) agreement with our
    previous Relativistic Effective Core Potential (RECP)-based~\cite{Titov} and semiempirical~\cite{Kozlov,Kozlov1}
    calculations.
 Recently a new technique to account for the most
important types of the core-valence correlations and the core polarization
effects with the help of the atomic Many Body Perturbation Theory~(MBPT)
was developed and proved to be very efficient for the calculations of the
hyperfine structure and $P,T$-odd effects for the BaF
molecule~\cite{Kozlov2}. Below we report the results of the application of
this method to calculation of the YbF molecule.
    The deviation of the calculated  $A_{\rm d}$  from the experimental value
    is analyzed and the final  $A_{\rm d}$  magnitude is corrected with the
    help of the semiempirical procedure~\cite{Kozlov1}.

\paragraph{Spin-rotational Hamiltonian.}

For the $^{171}$Yb isotope, which has nuclear spin $I=\case{1}{2}$, the
molecular spin-rota\-tional degrees of freedom are described by the
following spin-rotational Hamiltonian (see~\cite{KL}):
\begin{eqnarray}
        H_{\rm sr} & = & B{\bf N}^2 + \gamma {\bf S N} -D_e{\bf n E}
                 + {\bf S \hat{A} I}
\nonumber\\
                   & + & W_{\rm A} k_{\rm A} {\bf n \times S \cdot I}
                         +(W_{\rm S} k_{\rm S}
                         + W_d d_{\rm e}) {\bf S n}.
\label{1}
\end{eqnarray}
In this expression $\bf N$ is the rotational angular momentum, $B$ is the
rotational constant, $\bf S$ and $\bf I$ are the spins of the electron and
the Yb nucleus, $\bf n$ is the unit vector directed along the molecular
axis from Yb to F.  The spin-doubling constant $\gamma$ characterizes the
spin-rotational interaction. $D_e$ and $\bf E$ are the molecular dipole
moment and the external electric field.  The axial tensor $\bf \hat{A}$
describes magnetic hyperfine structure on the Yb nucleus.  It can be
determined by two parameters:  $A=(A_{\parallel}+2A_{\perp})/3$ and $A_{\rm
d}=(A_{\parallel}-A_{\perp})/3$. The smaller hyperfine structure
associated with the $^{19}$F nucleus is
 neglected.
 The last three
terms in \eref{1} account for the $P$- and $P,T$-odd effects.  First of
them describes interaction of the electron spin with the anapole moment of
the nucleus $k_{\rm A}$ \cite{FK}. The second one corresponds to the scalar
$P,T$-odd electron-nucleus interaction with the dimensionless constant
$k_{\rm S}$.  The third one describes interaction of the electron EDM
$d_{\rm e}$ with the internal molecular field $\bbox{E}^{\rm mol}$:

\begin{eqnarray}
        &&H_{d} = 2 d_{\rm e}
        \left(\begin{array}{cc}
        0 & 0 \\
        0 & \bbox{\sigma}
        \end{array} \right)
        \cdot \bbox{E}^{\rm mol},
\label{1a}\\
        &&W_d d_{\rm e} =
        2 \langle ^2\Sigma_{1/2}|H_{d}| ^2\Sigma_{1/2} \rangle .
\label{1b}
\end{eqnarray}
These equations show that the constant $\case{1}{2}W_d$ characterizes an
effective electric field on the unpaired electron.
    Similar to the $W_d$, the other $P$- and $P,T$-odd constants $W_i$
    depend on the electron spin-density in a vicinity of the heavy nucleus
    and reliability of their calculation can be also tested by comparison
    of the calculated and experimental hyperfine constants.

\paragraph{Calculation scheme.}

The Generalized RECP or
GRECP (see~\cite{GRECP} and references) calculation of the ground state
$^2\Sigma$ of YbF molecule was performed by analogy with~\cite{Titov}.  The
main difference of the present calculation is that the pseudospinors
corresponding to the $5s$- and $5p$-shells were frozen
from
 the calculation
of the Yb$^{2+}$ ion with the help of the level-shift technique (which is
also known as Huzinaga-type ECP, see e.g.~\cite{Huzinaga}). It was
necessary to freeze these pseudospinors, because polarization of the
corresponding shells was taken into account by means of the Effective
Operator (EO) technique (see~\cite{Kozlov2} for details).
RASSCF~\cite{RASSCF} calculations with 5284 configurations were performed
for 11 electrons distributed in RAS-1=(2,0,0,0), RAS-2=(2,1,1,0), and
RAS-3=(6,4,4,2) subspaces.
   The discribed above procedure follows the same lines as the one
   used in our previous calculation of BaF molecule~\cite{Kozlov2}.

 The molecular relativistic spinor for the unpaired electron was
 constructed from the molecular pseudoorbital $ \widetilde{\varphi}^M_u$
\begin{equation}
 \widetilde{\varphi}^M_u =
          \sum_i C^s_i \widetilde{\varphi}^s_i +
          \sum_i C^p_i \widetilde{\varphi}^{p,m_l=0}_i +
          \cdots,
 \label{3_1}
\end{equation}
 so that the atomic $s$- and $p$-pseudoorbitals of ytterbium in~(\ref{3_1})
 were replaced by the unsmoothed four-component DF spinors
 derived for the same atomic configurations which were used in generation of
 basis $s$- and $p$-pseudoorbitals.  The MO~LCAO coefficients were
preserved after the RECP calculations.  As the spin-orbit interaction for
the unpaired electron is small, the ``spin-averaged'' valence atomic
 $p$-pseudoorbital was replaced by the linear combination of the
 corresponding spinors with $j = l \pm 1/2$ (see~\cite{Titov} for details).

EOs for the magnetic hyperfine interaction, for the EDM
interaction~\eref{1a} and for the anapole moment interaction were
constructed by means of the atomic MBPT. The EOs include two most important
correlation corrections, which involve all the core electrons. The first
correction corresponds to the Random Phase Approximation (RPA), and the
second one corresponds to the substitution of the valence DF orbitals by
the Brueckner orbitals. The latter are found by solving the one-particle
equations with the self-energy operator $\sigma(\varepsilon)$ added to the
Dirac-Fock operator $h_{\rm DF}$:
\begin{equation}
        \left( h_{\rm DF}+\sigma(\varepsilon_n) \right) \phi_n
        = \varepsilon_n \phi_n.
\end{equation}
For the operator $\sigma(\varepsilon)$ we used diagrammatic technique
developed in \cite{DFK}. In our calculations of this operator we have
neglected excitations from the shells $1s \cdots 3d$. Opposite to that,
in RPA equations it was important to include all the core electrons.
Both RPA equations and Brueckner equations were solved for a finite
four-component basis set in the $V^{N-2}$ approximation (which means that
SCF corresponds to Yb$^{2+}$), and matrix elements of the EOs were
calculated. The basis set included DF orbitals for $1s \ldots 6s,6p,5d$
shells. In addition $7-18s,7-18p,6-17d,5-17f$ and $5-14g$ orbitals were
formed by analogy with the basis set N2 of~\cite{DFK}.  Molecular orbitals
were reexpanded in this basis set to find matrix elements of EOs for the
molecular wave function.

The main advantage of this method is that there is no need to extend the
active (valence) space in order to include core electrons. It is of
particular importance when one is interested in such molecular properties as
hyperfine, $P$-odd, and $P,T$-odd constants. Corresponding electronic
operators are singular at the nucleus. For this reason, it is necessary to
include all the core electrons in RPA equations. Corresponding extension of
the active space would be extremely expensive. Another approach which allows
core electrons to contribute to the spin density is the unrestricted DF
molecular calculation (where polarization is taken into
account~\cite{Parpia}).

\paragraph{Results.}

Explicit expression for the parameter $W_d$ is given in~\eref{1a} and
\eref{1b}. Other electronic matrix elements, which correspond to the
parameters $A$, $A_{\rm d}$ and $W_i$ of operator~\eref{1} can be found
in~\cite{KL}. All the radial integrals and atomic four-component spinors were
calculated for the finite nucleus $^{171}$Yb in a model of uniformly charged
ball. It is well known, that atomic matrix elements of operator~\eref{1a} are
proportional to $Z^3$.  The same scaling is applicable to the constant
$W_{\rm S}$, while the matrix elements which contribute to the constants
$W_{\rm A}$ are proportional to $Z^2$. As far as the nuclear charge of the
fluorine is  8 times smaller than that of the ytterbium, we have neglected
contributions to the $W_i$ parameters from the vicinity of the fluorine
nucleus. The additional argument which justifies this approximation is that
the unpaired electron in the YbF molecule is
 mainly
that of the Yb atom, and thus
the spin density is localized near the Yb atom. At present we do not have RPA
program for the scalar $P,T$-odd interaction, so we have focused here on the
calculation of the constants $W_d$ and $W_{\rm A}$.

Results for the parameters of the spin-rotational Hamiltonian are given in
table~\ref{P,T-odd}.  Comparison of the results of the GRECP/RASSCF
calculation \cite{Titov} with the results of the present GRECP/RASSCF/EO
calculation confirms our previous conclusion that correlations (including
spin polarization) with the core give a significant contribution to the
hyperfine structure and to the $P$-odd and $P,T$-odd constants.

Our final value for the hyperfine structure constant $A$ differs by less
than 3\% from the experimental value~\cite{Knight}. That means that the
symmetrical part of the spin density in the vicinity of the Yb nucleus in
our calculation is probably rather good. However, it is also very important
to reproduce the asymmetrical part of the spin density, which accounts for
the dipole constant $A_{\rm d}$. Our value for $A_d$ is in a slightly better
agreement with the experiment than the value from~\cite{Quiney}, but is
still underestimated by almost 23\%. About a half of this difference can be
explained by the fact that in our molecular calculation $4f$-shell of the Yb
atom was frozen. It was first pointed out by Khriplovich that in the YbF
molecule excitations from the $f$-shell can be important. In particular,
they can explain the small value of the spin-doubling constant $\gamma$
\cite{Sauer}.  It was shown recently that contribution of the $f$-shell
excitations to the spin density can give significant correction to the
constant $A_d$~\cite{Kozlov1}. Using equations (19) and (31) from the paper
\cite{Kozlov1} we obtain the following estimates for the $f$-shell
excitation contributions to $A$ and $A_{\rm d}$:
\begin{equation}
        \delta A \approx -3 \mbox{ MHz}, \qquad
        \delta A_d \approx 15 \mbox{ MHz}.
\label{3_2}
\end{equation}
Note that these corrections arise from the admixture to the molecular wave
function of the configuration with the hole in the $4f$~shell. The weight
of this configuration was estimated in \cite{Kozlov1} to be of the order of
4\%. Such an admixture is a purely molecular effect and is not accounted for
by the EO technique. So, we can conclude that $\delta A_{\rm d}$ can be
added to our value for $A_{\rm d}$, resulting in $A_{\rm d} \approx 94$~MHz,
that is in a much better agreement with the experimental value 102~MHz.

It is important that similar contribution of the $4f$-shell excitations to
the constant $W_d$ is strongly suppressed. Indeed, operator \eref{1a} mixes
$f$- and $d$-waves. For Yb atom, $4d$-shell is very deep and its mixture
with the $4f$-orbitals by the molecular field is very small, while
$5d$-shell is weekly bound and does not penetrate into the core region.
Similar contributions to other constants $W_i$ are negligible due to the
contact character of the corresponding interactions.

One can see that the values of the $W_d$ constant from the unrestricted DF
calculation~\cite{Parpia}, the most recent semiempirical
calculation~\cite{Kozlov1} and the present GRECP/RASSCF/EO calculation are in
a very close agreement now.
    It is also important that the valence electron contribution to the
    $W_d$ in~\cite{Parpia} is in 7.4\% agreement with the corresponding
    RECP-based calculation~\cite{Titov} (see table~\ref{P,T-odd}).
Another recent DF calculation~\cite{Quiney} gives the value, which is
two times smaller.

\paragraph*{Acknowledgments.}
This work was implemented under the financial support of the Russian
Foundation for Basic Research (N.~M.\ and A.~T.: grants N 96--03--33036
and 96--03--00069g; M.~K.: grant N 98--02--17663).

\newpage
\mediumtext

\begin{table}
\caption{Parameters of the spin-rotational Hamiltonian for $^{171}$YbF.}
\begin{tabular}{lcrdrr}
                        &   $A$   &$A_{\rm d}$& $W_d$    & $W_{\rm A}$
                        & $W_{\rm S}$ \\
Method                  & (MHz)   &  (MHz)
                        & ($10^{25}~\frac{\rm Hz}{\rm e\cdot cm}$)
                                                         & (Hz)  & (kHz) \\
\hline
Exper.\cite{Knight}     &  7617   &  102      &          &       &       \\
Semiemp.\cite{Kozlov}   &         &           & $-$1.5   &  730  & $-$48 \\
GRECP/SCF\cite{Titov}   &  4932   &   59      & $-$0.91  &  484  & $-$33 \\
GRECP/RASSCF\cite{Titov}&  4854   &   60      & $-$0.91  &  486  & $-$33 \\
Semiemp.\cite{Kozlov1}  &         &           & $-$1.26  &       & $-$43 \\
DHF\cite{Quiney}        &  5918   &   35      & $-$0.31  &  163  & $-$11 \\
DHF+CP\cite{Quiney}     &  7865   &   60      & $-$0.60  &  310  & $-$21 \\
DHF (rescaled)
\cite{Quiney}           &         &           & $-$0.62  &  326  & $-$22 \\
DF (unpaired elect-     &         &           &          &       &       \\
~~~ron contribution)
\cite{Parpia}           &         &           & $-$0.962 &       &       \\
Unrestricted DF\cite{Parpia} &    &           & $-$1.203 &       & $-$22 \\
GRECP/RASSCF/EO         &         &           &          &       &       \\
~~~(this work)          &  7842   &   79      & $-$1.206 &  634  &       \\
GRECP/RASSCF/EO         &         &           &          &       &       \\
~~~(with $4f$-hole
   correction)          &  7839   &   94      & $-$1.206 &  634  &
\end{tabular}
\label{P,T-odd}
\end{table}


\begin{references}
\bibitem{Kozlov}     M.G.Kozlov, V.F.Ezhov,
                Phys.Rev.{\bf A}, {\bf 49}, 4502 (1994).
\bibitem{Titov}      A.V.Titov, N.S.Mosyagin, V.F.Ezhov,
                Phys.Rev.Lett., {\bf 77}, 5346 (1996).
\bibitem{Kozlov1}    M.G.Kozlov,
                J.Phys.{\bf B}, {\bf 30}, L607 (1997).
\bibitem{Quiney}     H.M.Quiney, H.Skaane, I.P.Grant,
                J.Phys.{\bf B}, {\bf 31}, L85 (1998).
\bibitem{Parpia}     F.A.Parpia,
                J.Phys.{\bf B}, {\bf 31}, 1409 (1998).
\bibitem{Hinds}      B.E.Sauer, S.K.Peck, G.Redgrave, E.Hinds,
                Phys.Scr., {\bf T70}, 34 (1997).
\bibitem{KL}         M.G.Kozlov, L.N.Labzowsky,
                J.Phys.{\bf B}, {\bf 28}, 1933 (1995).
\bibitem{Kozlov2}    M.G.Kozlov, A.V.Titov, N.S.Mosyagin, P.V.Souchko,
                Phys.Rev.{\bf A}, {\bf 56}, R3326 (1997).
\bibitem{FK}         V.V.Flambaum, I.B.Khriplovich,
                Phys.Lett., {\bf 110A}, 12 (1985).
\bibitem{GRECP}      N.~S.~Mosyagin, A.~V.~Titov, and Z.~Latajka,
                Int.\ J.\ Quant.\ Chem.\ {\bf 63}, 1107 (1997);
                        A.~V.~Titov and N.~S.~Mosyagin,
                Preprint PNPI No.\ {\bf 2182} (Petersburg Nuclear
                Physics Institute, Gatchina, St.-Petersburg district, 1997),
                81~p., submitted for publication.
\bibitem{Huzinaga}   V.Bonifacic, S.Huzinaga,
                J.Chem.Phys., {\bf 60}, 2779 (1974).
\bibitem{RASSCF}     J.Olsen, B.O.Roos,
                 J.Chem.Phys., {\bf 89}, 2185 (1988);
                     K.~Andersson, M.~R.~A.~Blomberg,
                M.~P.~F\"{u}lscher, V.~Kell\"{o}, R.~Lindh,
                P.-Aa.~Malmqvist, J.~Noga, J.~Olsen, B.~O.~Roos,
                A.~J.~Sadlej, P.~E.~M.~Siegbahn, M.~Urban, P.-O.~Widmark,
                MOLCAS, Version 2, University of Lund, Sweden 1991.
\bibitem{DFK}   V.A.Dzuba, V.V.Flambaum, M.G.Kozlov,
                JETP Lett., {\bf 63}, 882 (1996);
                Phys.Rev.{\bf A}, {\bf 54}, 3948 (1996).
\bibitem{Knight}
        L. B. Knight, Jr. and W. Weltner, Jr., J. of Chem. Phys.
        {\bf 53}, 4111 (1970).
\bibitem{Sauer}     B.E.Sauer, J.Wang, E.A.Hinds,
                Phys.Rev.Lett., {\bf 74}, 1554 (1995).
\end{references}
\end{document}